\documentstyle[12pt,epsf,aps,preprint]{revtex}

\newcommand{\be}{\begin{equation}}
\newcommand{\ee}{\end{equation}}
\newcommand{\ba}{\begin{array}{c}}
\newcommand{\ea}{\end{array}}
\newcommand{\bqa}{\begin{eqnarray}}
\newcommand{\eqa}{\end{eqnarray}}

\begin{document}
\tightenlines \draft
\title{Top quark, heavy fermions and the composite Higgs boson}
\author{Bin Zhang and Hanqing Zheng}
\address{Department of Physics, Peking University, Beijing 100871,\\
People's Republic of China} \maketitle

\begin{abstract}
We study the properties of heavy fermions in the vector-like
representation of the electro-weak gauge group $SU(2)_W\times
U(1)_Y$ with Yukawa couplings to the standard model Higgs boson.
Applying the renormalization group analysis, we discuss the
effects of  heavy fermions to the vacuum stability bound and the
triviality bound on the mass of the Higgs boson. We also discuss
the interesting possibility that the Higgs particle is composed of
the top quark and heavy fermions. The bound on the composite Higgs
mass is estimated using the method of Bardeen, Hill and
Lindner~\cite{BHL91}, $150$GeV$\leq m_H\leq$ $450$GeV.
\end{abstract}


Enormous efforts have been made in searching for physics beyond
the standard model but up to now a crucial, direct experimental
indication is still illusive. One of the most important motivation
to study the property of heavy fermions above the energy scale
accessible by current accelerators is to look for extra building
blocks of nature beyond the three families of the standard model.
For this purpose it may be adequate to study fermions in
vector-like representations of the electro-weak gauge group with a
large bare mass term, rather than the conventional chiral
fermions. The main reason for this is from the strong experimental
constraints on the S parameter~\cite{peskin}. While experiments
favor a negative value of S~\cite{exp}, a standard chiral doublet
of heavy fermions (degenerate in mass) contributes to the S
parameter as $1/6\pi $. On the contrary, for fermions in the
vector-like representation of the electro-weak gauge group, a
large bare fermion mass M completely changes the low energy
properties of the heavy fermions. As a consequence of the
decoupling theorem, heavy fermions' contribution to the oblique
corrections of the standard processes are suppressed by
${\frac{1}{M^2}}$. Especially, their contribution to the S
parameter is still positive definite but much smaller in magnitude
than the ordinary chiral fermions. Furthermore the heavy fermion
contributes to the vacuum expectation value of electroweak
symmetry breaking as~{\cite{zheng}},
\begin{equation}
\delta (f_\pi ^2)=\delta v^2\simeq {\frac{m^2N_c}{2\pi ^2}}(\log
{\frac{ \Lambda ^2}{M^2}}),  \label{dv}
\end{equation}
where where $m$ is the mass generated by the Yukawa coupling and
$\Lambda $ is the cutoff scale of the effective theory. It is
interesting to compare the above expression to that of the pion
decay constant obtained in the QCD effective action
approach~\cite{rafael}, $f_\pi^2={\frac{N_c}{4\pi ^2}}M_Q^2\ln
({\frac{\Lambda _{QCD}^2}{M_Q^2}})$, where $ M_Q$ is the
constitute quark mass which is similar to $m$ in our present
discussion. We notice that if in the above Eq.~({\ref{dv}) $m\sim
O(v)$ then several of these heavy fermions would be enough to
induce the electroweak symmetry breaking. Therefore if there is a
strong attractive forces in the appropriate channel to cause the
heavy fermion condensation then they may place the role similar to
techniquarks in the technicolor model. This way of dynamical
electroweak symmetry breaking, if possible, is remarkable.
Contrary to the technicolor model, it avoids the dangerous low
energy consequences which may contradict experiments.} Also it can
be  demonstrated~\cite{vhiggs} that the composite Higgs boson's
mass is proportional to the dynamically generated fermion mass and
completely decouples from the bare one, even though the Higgs
particle is ``composed of'' the heavy fermions. This is a
consequence of  symmetry and be model independent, at least in a
system with second order phase transition.

{\ }Heavy fermions may have many other interesting role in physics beyond
the standard model either. For example, they may be responsible for a
dynamical generation of light fermion mass matrix~\cite{nielsen}; they
appear in the ``vector--like extension'' of the standard model\cite{fuji};
they are natural consequences of many grand unification models, and of the
super-symmetric preon model~\cite{pati}. Therefore it is important to
investigate the fundamental properties of the heavy vector-like fermions
thoroughly.

There have been continuous interests in understanding the
structure of the standard model at high energies, even up to
Planck scale~(see for example, \cite{stab,CEQ,vac,HS,kang} and the
most recent review which contains many materials,
Ref.~\cite{fhs}).  A powerful tool is to use the renormalization
group equations to trace the evolution of the coupling constant of
the $\lambda \phi ^4$ self-interaction of the Higgs particle.
Assuming the standard model remains valid up to certain scale
$\Lambda $, an upper bound (the triviality bound, obtained by
requiring $\lambda $ not to blow up below $\Lambda $) of the Higgs
boson mass, $m_H$, can be obtained. Meanwhile, requiring the
stability of the electro-weak vacuum, we can also obtain a lower
bound on $m_h$. For the later purpose, in principle one needs to
consider the renormalization group improved effective
potential~\cite{sher} and require it be bounded from below. But in
practice this turns out to be equivalent to the requirement that
the Higgs self-interaction coupling constant $\lambda $ does not
become negative, below the given scale (see \cite{stab} and ref.
therein). It is remarkable that for the given experimental value
of the top quark mass (here we use $m_t=174$GeV), there is an
allowed range for the Higgs boson mass, $130\hbox {GeV}\leq
$$m_H$$\leq 200\hbox{GeV}$~\cite{stab}, for which the standard
model may remain valid up to Planck scale.

In this paper we devote to study heavy fermions' influence to the
vacuum stability bound and the triviality bound on the Higgs boson
mass. Furthermore, assuming that the Higgs boson is a composite
particle, we use the method developed in Ref.~\cite{BHL91} to
estimate the range of the Higgs boson's mass\footnote{ This paper
replaces and is an extension of  Ref.~\cite{heaven}.}. We find
that the top quark also place an important role in the
compositeness picture and the composite Higgs boson can be viewed
as a mixture of $\bar tt$ pair and heavy fermion pair. The larger
the hierarchy is the more top quark content the composite Higgs
boson contains and vise--versa.

We start with the following general Lagrangian for heavy fermions,
\[
{\cal L}=\bar Q(i{\Delta \llap{$/$}}_d-M)Q+\bar U(i{\Delta \llap{$/$}}%
_s-M)U+\bar D(i{\Delta \llap{$/$}}_s-M)D+g_d\bar Q_L\phi D_R+g_u\bar
Q_L\tilde \phi U_R
\]
\begin{equation}
+g_d^{\prime }\bar Q_R\phi D_L+g_u^{\prime }\bar Q_R\tilde \phi U_L+h.c.\ .
\label{L2}
\end{equation}
In above $Q$ is the $SU(2)_W$ doublet and $U$ and $D$ are singlets with weak
hypercharge $Y_Q$, $Y_U$ and $Y_D$, respectively (with the selection rule $%
Y_U-Y_Q=Y_Q-Y_D=Y_\phi $). We assume they participate in strong
interactions and are in fundamental representations of $SU(3)_C$.
The subscript $d$ ($s$) in the covariant derivatives denotes that
the corresponding fermion is a $ \hbox{SU(2)}_W$ doublet (singlet)
and $\phi $ denotes the standard Higgs doublet. We further expect
the Yukawa couplings to be of order 1. For simplicity we take all
the bare fermion masses to be equal. Also we do not discuss the
mixing between heavy fermions and the ordinary fermions here.

As is well known, because of the negative sign, fermions turn to
destabilize the vacuum. After including heavy fermions the
structure of our world changes drastically at high scales, even
though vector-like fermions are essentially decoupling below their
threshold. At scales much higher than the threshold whether the
fermion field is chiral or vector-like does not make any
qualitative difference. The only thing matters is the number of
independent Yukawa couplings and their strength. The relevant one
loop RGEs are listed as below\footnote{ Due to a careless mistake,
the Yukawa coupling RGEs given in Ref.~\cite {heaven} contain an
error. The top quark effects were not considered correctly.},

\begin{equation}  \label{l}
16\pi^2{\frac{d\lambda}{dt}}= 24\lambda^2+12\lambda A
-6A^{\prime}-(9g_2^2+3g_1^2)\lambda \\
+{\frac{9}{8}}g_2^4+{\frac{3}{4}}g_2^2g_1^2 +{\frac{3}{8}}g_1^4\ ,
\end{equation}

\begin{equation}  \label{rgeh}
16\pi^2{\frac{dg_u}{dt}}= \{ {\frac{3}{2}}(g_ug_u^\dagger-g_dg_d^\dagger)+3A
-8g_s^2-{\frac{9}{4}}g_2^2-3(Y_Q^2+Y_U^2)g_1^2\}g_u\ ,
\end{equation}

\begin{equation}
16\pi^2{\frac{dg_d}{dt}}= \{ {\frac{3}{2}}(g_dg_d^\dagger-g_ug_u^\dagger)+3A
-8g_s^2-{\frac{9}{4}}g_2^2-3(Y_Q^2+Y_D^2)g_1^2\}g_d\ ,
\end{equation}

\begin{equation}  \label{rget}
16\pi^2{\frac{dg_t}{dt}}= \{ {\frac{3}{2}}g_t^2+3A
-8g_s^2-{\frac{9}{4}} g_2^2-{\frac{17}{12}}g_1^2\}g_t\ ,
\end{equation}
where,
\begin{equation}
A=\hbox{tr} \{g_ug_u^\dagger+g_dg_d^\dagger+ g^{\prime}_u
(g^{\prime}_u)^\dagger+g^{\prime}_d(g^{\prime}_d)^\dagger\} +g_t^2\ ,
\end{equation}
\begin{equation}
A^{\prime}=\hbox{tr} \{ (g_ug_u^\dagger)^2+(g_dg_d^\dagger)^2+
(g^{\prime}_u(g^{\prime}_u)^\dagger)^2+(g^{\prime}_d(g^{\prime}_d)^%
\dagger)^2\}+g_t^4\ ,
\end{equation}
and
\begin{equation}
16\pi^2{\frac{dg_s}{dt}}= (-7+ {\frac{2}{3}}(2N_Q+N_U+N_D)\theta )g_s^3\ ,
\end{equation}
\begin{equation}
16\pi^2{\frac{dg_2}{dt}}= (-{\frac{19}{6}} +2N_Q\theta)g_2^3\ ,
\end{equation}
\begin{equation}
16\pi^2{\frac{dg_1}{dt}}= ({\frac{41}{6}}+ 4(2N_QY_Q^2+N_UY_U^2+N_DY_D^2
)\theta)g_1^3\ ,
\end{equation}
where the trace doesn't sum over color space and $g^{\prime}_u$
and $g^{\prime}_d$ obey similar equations. In general these Yukawa
couplings can be matrices in the flavor space if there are many
heavy fermions, and $g_t$ is the Yukawa coupling of the top quark
($g_t=\sqrt{2}m_t/v$). The symbols $N_Q$, $N_U$ and $N_D$ refer to
the number of Q, U and D type of quarks, respectively. We use a
simple step function $\theta=\theta(t-log(M/M_z))$ to model the
heavy fermion threshold effects. All the Yukawa couplings in above
renormalization group equations are understood as multiplied by
$\theta$.  Applications using two loop RGEs in the standard model
case and beyond was considered in Ref.~\cite{PZ} and it was found
that the two loop effects are very small below Planck scale.

\begin{figure}[hbtp]
\begin{center}
\vspace*{-20mm} \mbox{\epsfysize=70mm\epsffile{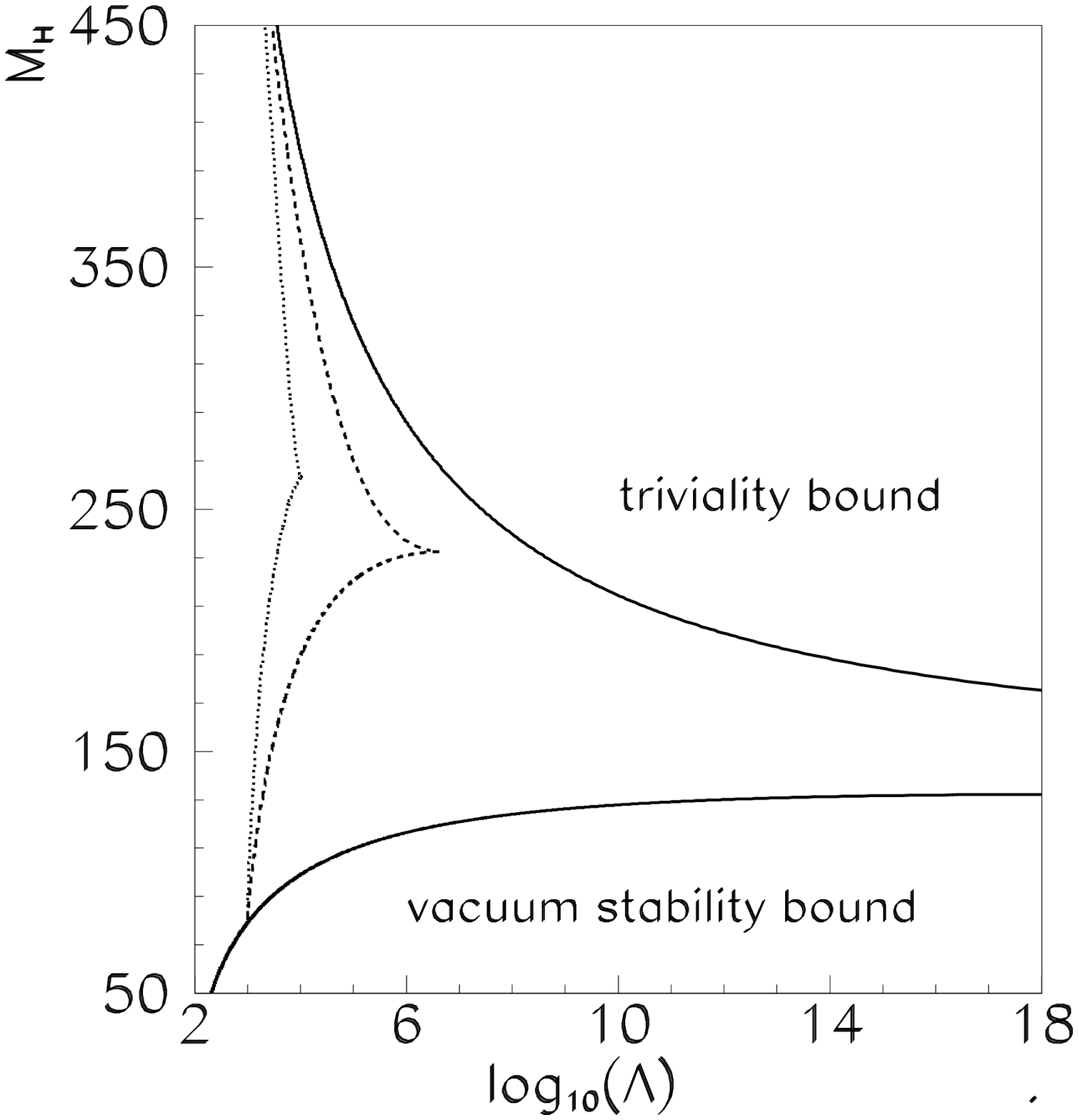}}
\vspace*{0mm} \caption{\label{fig1}
 Vacuum stability and
triviality bounds on the Higgs mass as a function of $\Lambda$.
The solid lines are the standard model case The dashed (dotted)
lines correspond to $N=1$ ($N=3$), the Yukawa coupling $g_Y=1$.}
\end{center}
\end{figure}

In the following qualitative discussion, we set $Y_Q=1/6$, $Y_U=2/3$ and $%
Y_D=-1/3$. For simplicity we take $N_Q=N_U=N_D$ ($\equiv N$) and
all the Yukawa couplings (after the diagonalization of the
coupling matrices) in the initial boundary conditions being
identical~\footnote{The `up' and `down' type quarks evolve
differently because of different $ U(1)_Y$ charge, however the
isospin splitting is very small for the standard values of the
hypercharge.}. In fig.~\ref{fig1} we plot the vacuum stability
bound and the triviality bound on the Higgs mass as a function of
the scale $\Lambda $ for some typical values of the parameters of
the heavy fermions. We see that the inclusion of heavy fermions
drastically change the Standard model structure at high energies
even though they decouple from the low energy world. They tighten
the bound on the mass of the Higgs boson as a function of the
cutoff scale $\Lambda $. Notice that (in terms of one loop
renormalization equations) the upper line (triviality bound) and
the lower line (vacuum stability bound) never meet each other.
Because the upper line is drawn by requiring $\lambda $ not to
blow up and the lower line is drawn by requiring $\lambda \geq 0$.
Between them is the ultra-violet unstable fixed point of $\lambda
$, so the two lines get close to each other rapidly.

We now study the interesting possibility of considering the Higgs particle
as a composite object of the heavy vector-like fermions. Applying the above
renormalization group analysis to the composite model leads to some
interesting results which we present below. We follow the method proposed by
Bardeen, Hill and Lindner (BHL) \cite{BHL91} originally developed for the
top quark condensate model. The basic idea of the BHL method is the
following: Using the collective field method the four--fermi interaction
Lagrangian can be rewritten into an effective Higgs--Yukawa interaction
Lagrangian at the cutoff scale $\Lambda $. The effective Yukawa interaction
Lagrangian is identical to the standard model at the cutoff scale $\Lambda $%
, but with vanishing wave function renormalization constant of the
Higgs field ($Z_H=0$) and vanishing Higgs self-coupling ($\lambda
=0$). Below $ \Lambda $ the model is equivalent to the standard
model and therefore the coupling constants of the effective theory
run according to the standard model renormalization group
equations. However the vanishing of $Z_H$ at the scale $\mu
=\Lambda $ leads to the following boundary conditions of the
renormalization group equations:
\begin{equation}
g_Y^r\to \infty \ ,\,\,\,\,\lambda ^r/(g_Y^r)^4\to 0\ ,
\end{equation}
where $\lambda ^r$ and $g_Y^r$ are the renormalized Higgs self-coupling and
Yukawa coupling, respectively. With the renormalization group equations and
boundary conditions, one can predict the mass of the Higgs boson and the
fermion mass (or the Yukawa couplings) at the infra-red fixed point. In the
present case, of course, the ``standard model'' often refers to the standard
model plus heavy fermions and the ``infra-red fixed point'' value of $g_Y$
refers to its value at the threshold.

\begin{figure}[hbtp]
\begin{center}
\vspace*{-10mm} \mbox{\epsfysize=70mm\epsffile{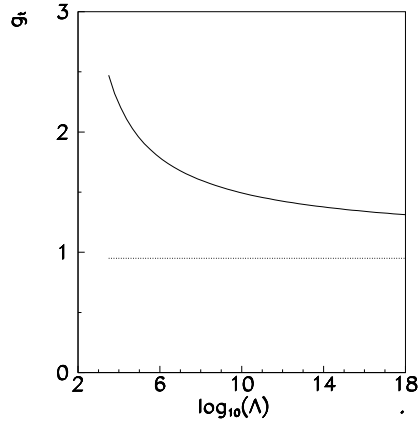}}
\vspace*{0mm} \caption{ \label{fig2} The solid line: Infra-red
fixed point value of $g_t$ as a function of the compositeness
scale according to the standard model RGEs. The dotted line
indicates the experimental value of $g_t$. }
\end{center}
\end{figure}

The minimal top quark condensate model has already been ruled out by
experiments. In order to generate the electroweak symmetry breaking scale $v$%
, the top quark mass is required to be at least as large as
218~GeV (corresponding to $\Lambda =10^{19}$~GeV, i.e., Planck
scale). The experimental value of the top quark mass indicates
that the top quark Yukawa coupling does not diverge up to Planck
scale in the standard model and therefore does not meet the
compositeness condition of BHL. This can be clearly seen from
fig.~\ref{fig2}. However, in the present model, since there is no
strict experimental constraint on the heavy fermions, the
compositeness condition is easily and naturally achievable, that
$g_t$ blows up below Planck scale with the aid of the heavy
fermions. From Eqs.~(\ref{rgeh}), (\ref{rget}) we see that the
evolution of the Yukawa couplings are correlated to each other and
one `blows up' leads the another to blow up too.

\begin{figure}[hbtp]
\begin{center}
\vspace*{-10mm} \mbox{\epsfysize=70mm\epsffile{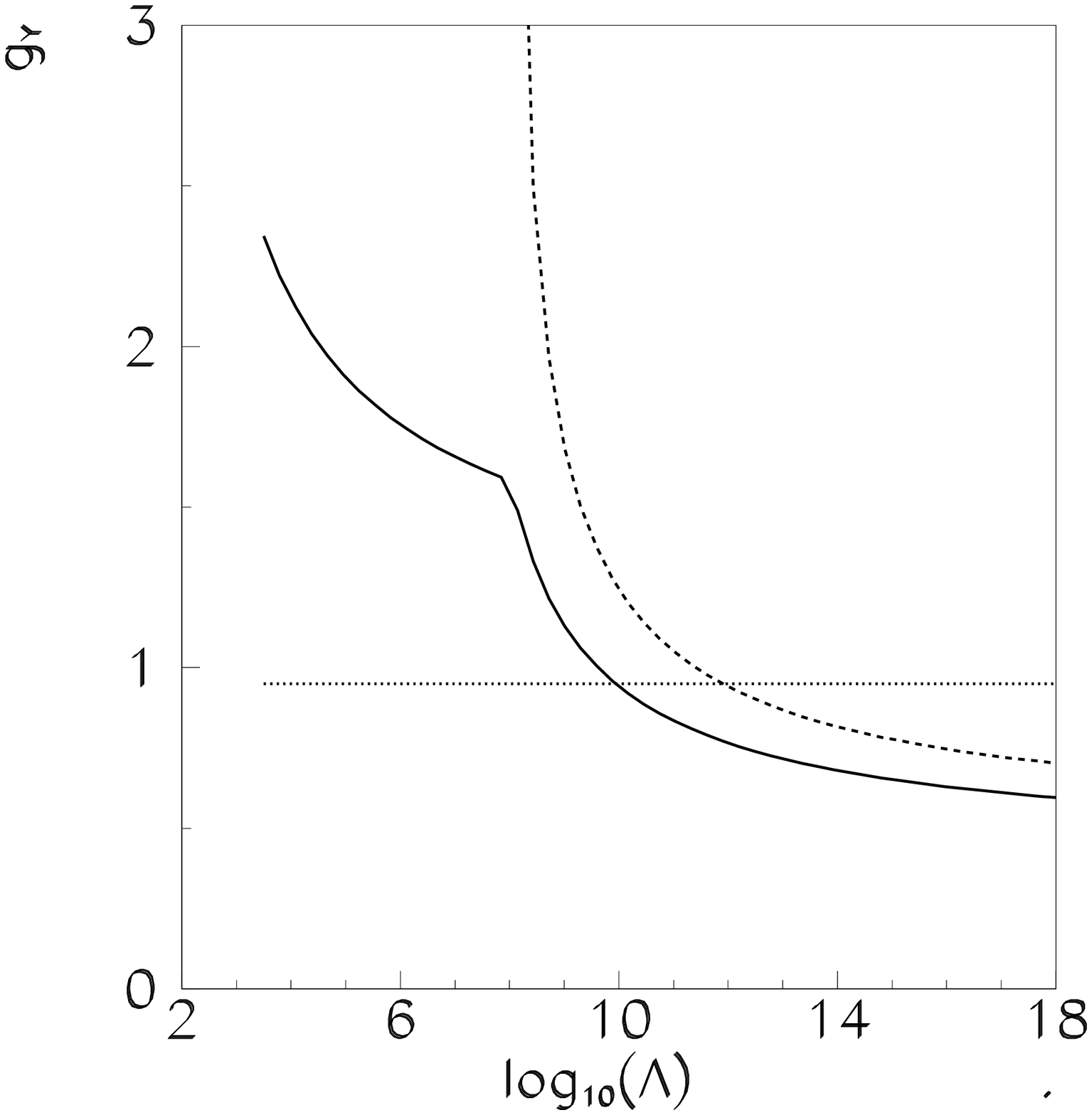}}
\caption{ \label{fig3}%
Infra-red fixed
point value of $g_t$ (solid line) and $g_Y$ (dashed line).
The dotted line
indicates $g_t^{exp}$. M=$10^8$~GeV, N=1. The correct value of the
composite scale is at where the solid line cross the dotted line.}
\end{center}
\end{figure}

\begin{figure}[hbtp]
\begin{center}
\vspace*{-10mm} \mbox{\epsfysize=70mm\epsffile{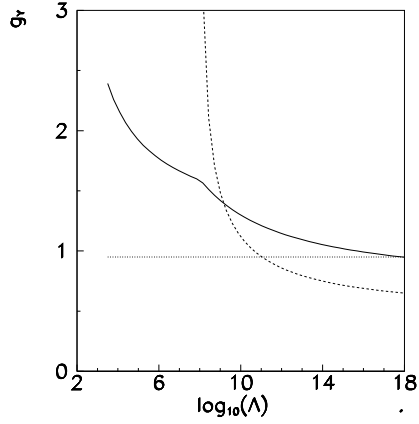}}
\vspace*{10mm}
\caption{ \label{fig4}%
Infra-red fixed
point value of $g_t$ (solid line) and $g_Y$ (dashed line).
The dotted line
indicates $g_t^{exp}$. M=$10^8$~GeV, N=1.
Here there are more top quark content in the composite Higgs boson.
}
\end{center}
\end{figure}

When both the top quark and heavy fermions are involved, the
situation is more complicated than the simple top condensate
model. Running the RGEs down from certain scale, one must take
good care of $g_t$ to ensure that it reaches the experimental
value at the infra-red fixed point. This means that a certain
fine-tuning is needed on the initial boundary conditions of the
Yukawa coupling RGEs. The composite Higgs boson is now a mixture
of $\bar tt$ pairs and the heavy quark pairs. Fig.~\ref{fig3} and
fig.~\ref{fig4} show two typical examples of such a situation. In
the situation of fig.~\ref{fig3} the Higgs particle is mainly
composed of heavy fermions while in fig.~\ref {fig4} the top quark
becomes the major component. Notice that for a given ratio of
$g_Y/g_t$ in the compositeness boundary condition (for fixed M and
N), the composite scale $\Lambda $ is no longer free, rather it is
determined by $g_t^{exp}$.

\begin{figure}[hbtp]
\begin{center}
\vspace*{-10mm} \mbox{\epsfysize=70mm\epsffile{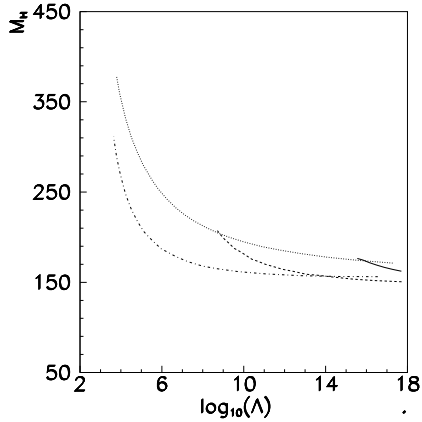}}
\vspace*{10mm}
\caption{ \label{fig5}%
IR fixed point
value (at $M_Z$) of $M_H$
as a function of
the compositeness scale.
The solid line: $N=3$, $M=10^{15}$~GeV;
the dashed  line: $N=3$, $M=10^8$~GeV;
the dotted line: $N=1$, $M=10^3$~GeV;
the dot-dashed line: $N=3$, $M=10^3$~GeV. }
\end{center}
\end{figure}

In fig.~\ref{fig5} we plot the composite Higgs particle's
mass\footnote{ The Higgs mass in these figures is the renormalized
mass at $\mu=M_Z$. The renormalized mass is close to the pole mass
of the Higgs boson.} as a function of the composite scale,
$\Lambda $. We chose $N\leq 3$ to avoid the problem with the
non-asymptotic freedom of $g_s$. From fig.~\ref{fig5} we see that
the allowed range for the Higgs mass is rather narrow against the
wide range of the cutoff scale, the bare fermion mass and the
number of heavy fermions, except when the heavy fermion bare mass
$M$ is close to the cutoff $\Lambda $. A lower bound on the Higgs
mass can be obtained: $m_H\geq 150$~GeV. When $M$ is getting close
to the cutoff scale our results become unstable and are sensitive
to the input numerical values of the boundary conditions. In such
a situation the scale is not large enough for the couplings to
reach the infra-red stable point. It is estimated that the Higgs
mass will not exceed 450~GeV, otherwise the whole mechanism become
unnatural (in the sense that the Yukawa coupling constant at
electroweak scale also becomes substantially larger than 1).

\begin{figure}[hbtp]
\begin{center}
\vspace*{-10mm} \mbox{\epsfysize=70mm\epsffile{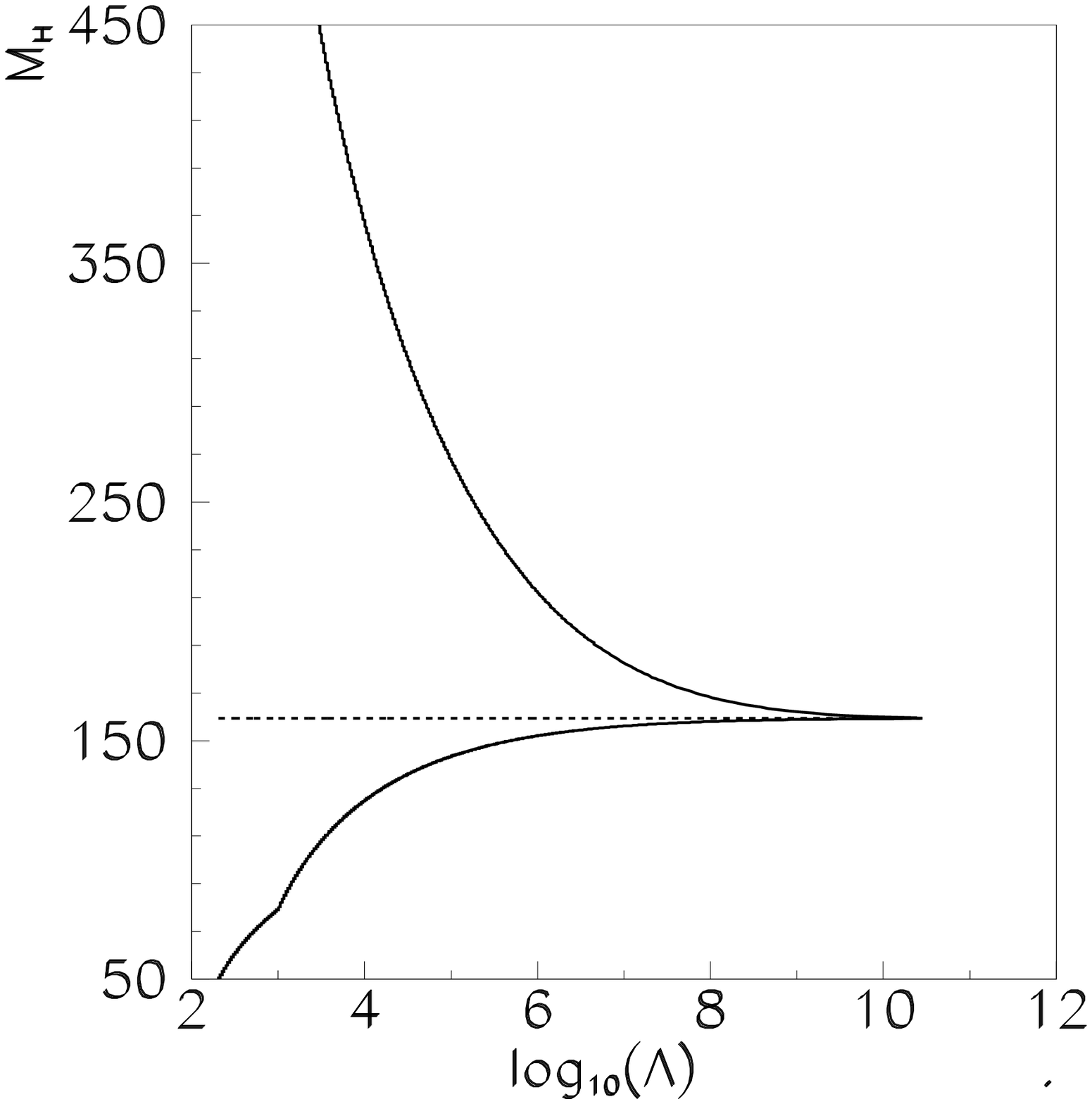}}
\vspace*{10mm} \caption{ \label{fig6}
 IR--UV fixed point structure
and compositeness. N=3, M=$10^3$~GeV, $\Lambda_C=10^{11}$.}
\end{center}
\end{figure}

In fig.~\ref{fig6} we plot a typical example of the Higgs mass for
a given cutoff scale $\Lambda _c$ and $N$. We also plot the
triviality bound and the vacuum stability bound using the value of
the Yukawa coupling constant at the infrared fixed-point, which is
determined uniquely by the parameters $M$, $\Lambda _c$ and $N$ in
the compositeness picture, as the initial boundary condition. It
is very interesting to notice that $m_H$ and $\Lambda $ take the
values where the curves of triviality bound and vacuum stability
bound (practically) meet each other. This is the unique feature of
BHL compositeness picture. The reason behind this is very simple:
The infra-red attractive fixed point corresponds to the
ultra-violet unstable fixed point. In the sense of
Ref.~\cite{hasen}, this picture can be disturbed. However in most
cases the infra-red--ultra-violet fixed point structure is
influential and rather stable against perturbation.

In above we presented an analysis on the properties of heavy
fermions in vector-like representations of the standard model
gauge group. We pointed out earlier~\cite{vhiggs} that if they can
place the role to break the electro-weak symmetry dynamically the
theory has some distinguishable properties: the low energy theory
is asymptotically renormalizable and returns to the standard
model. From the above RG analysis we realize that the top quark
also places an important role in the dynamical symmetry breaking
scenario and our model can be viewed as a natural generalization
to the top condensate model of BHL. We found that the composite
Higgs boson's mass ranges from 150GeV to 450GeV, and the lighter
the Higgs boson is the more top quark content it contains, and
vice versa. Our prediction to the mass of the Higgs boson will be
testable by  LHC and the model will be ruled out if $m_H$ is found
to be below 150GeV.

{\it Acknowledgment}: The work of H.Z.
 is supported in part by the National
Natural Science Foundation of China under grant No.~19775005.

\begin{thebibliography}{99}
\bibitem{BHL91}W.~A.~Bardeen, C.~T.~Hill and M.~Lindner, Phys. Rev. {\bf
D41} (1990) 1647.

\bibitem{peskin}M.~E.~Peskin and T.~Takeuchi, Phys. Rev. {\bf D46} (1992)
381.

\bibitem{exp}  for experimental constraints on the S parameter, see
for example, K.~Hagiwara, {\it Talk presented at XVII International
Symposium on Lepton and Photon Interactions at High Energies, 10-15 August
1995, Beijing, China}, Preprint KEK-TH-95-184.

\bibitem{zheng}  H.~Zheng, Phys. Rev. {\bf D51} (1995)251.

\bibitem{rafael} D.~Espriu, E.~de Rafael and J.~Taron, Nucl. Phys. {\bf B345
} (1990) 22.

\bibitem{vhiggs}  H.~Zheng, Phys. Rev. {\bf D52} (1995) 6500.

\bibitem{nielsen}  C.~D.~Froggatt and H.~B.~Nielsen, Nucl. Phys. {\bf B147}
(1979) 277.

\bibitem{fuji}  K.~Fujikawa, Prog. Theor. Phys. {\bf 92}, 1149(1994).

\bibitem{pati}  J.~Pati, Phys. Lett. {\bf B228}, 228(1989).

\bibitem{stab}  G.~Altarelli and G.~Isidori, Phys. Lett. {\bf B337} (1994)
141.

\bibitem{CEQ}  J.~A.~ Casas, J.~R.~Espinosa and M. Quiros, Phys. Lett {\bf %
B342} (1995) 171.

\bibitem{vac}  M.~A.~Diaz, T.~A.~ter Veldhuis and T.~J.Weiler, Phys. Rev.
Lett. {\bf 74} (1995) 2876.

\bibitem{HS}  P.~Q.~Hung and M.~ Sher, Phys. Lett. {\bf B374} (1996) 138.

\bibitem{kang}D. Dooling, K. Kang and S. K. Kang, preprint hep-ph9710258.
\bibitem{fhs}P.~Frampton, P.~Q.~Hung and M.~Sher, hep-ph/9903387.
\bibitem{sher}M.~ Sher, Phys. Rep. {\bf 179} (1989) 273.
\bibitem{heaven}  H.~Zheng, Phys. Lett. {\bf B370} (1996) 201;
Erratum {\bf B382} (1996) 448.
\bibitem{PZ}Yu.~F.~Pirogov, O.~V.~Zenin, Eur.~Phys.~J. {\bf
C10} (1999) 629.
\bibitem{hasen}  A.~Hasenfratz, P.~Hasenfratz, K.~Jansen, J.~Kuti and
Y.~Shen, Nucl. Phys. {\bf B365}, 79(1991); J. Zinn-Justin, Nucl.
Phys. {\bf B367}, 105(1991).
\end{thebibliography}
\end{document}